\documentstyle[]{mn}

\input{psfig} 

\def\hmpc {h^{-1} {\rm Mpc}}

\def\and  {{\it {et al.} }}

\newcommand{\etal}{{\it et al.\ }}
\newcommand{\xib}{\bar{\xi}}

\newcommand{\avg}[1]{\langle{#1}\rangle}

\newcommand{\ltsima}{$\; \buildrel < \over \sim \;$}
\newcommand{\lsim}{\lower.5ex\hbox{\ltsima}}
\newcommand{\gtsima}{$\; \buildrel > \over \sim \;$}
\newcommand{\gsim}{\lower.5ex\hbox{\gtsima}}

\begin{document}

\title[The luminosity dependence of clustering and higher order correlations in the PSCz survey]{The luminosity dependence of clustering and higher order correlations in the PSCz survey}

\author[I.Szapudi]{Istv\'an Szapudi$^{1,2}$, Enzo Branchini$^3$, 
        C. S. Frenk$^1$, Steve Maddox$^4$,  Will Saunders$^5$ 
%\altaffilmark{1}}
%\affil{Fermi National Accelerator Laboratory}
%\affil{Theoretical Astrophysics Group}
%\affil{Batavia, IL 60510}
%\altaffiltext{1}{E-mail:\ szapudi@astro1.fnal.gov}
\\
\\
$^1$ Department of Physics, University of Durham, South Road, 
Durham DH1 3LE, UK \\
$^2$ CITA, University of Toronto, 60 George Street, 
Ontario, Canada, M5S 3H8 (present address)\\
$^3$ Kapteyn Institute, University of Groningen, Landleven 12, 
Groningen, 9700 AV, the Netherlands\\
$^4$ School of Physics and Astronomy, University of Nottingham,
Nottingham NG7 2RD 
$^5$ Institute for Astronomy, Blackford Hill, Edinburgh EH9 9RJ, UK
\\
}

\voffset -0.5cm
\maketitle 
 
\begin{abstract}

We investigate the spatial clustering of galaxies in the PSCz galaxy
redshift survey, as revealed by the two-point correlation function,
the luminosity mark correlations, and the moments of
counts-in-cells. We construct volume-limited subsamples at different
depths, and search for a luminosity dependence of the clustering
pattern. We find no statistically significant effect in either the
two-point correlation function or the mark correlations and so we take
each subsample (of different characteristic luminosity) as
representing the same statistical process. We then carry out
a counts-in-cells analysis of the volume-limited subsamples, including
a rigorous error calculation based on the recent theory of Szapudi,
Colombi and Bernardeau. In this way, we derive the best estimates to
date of the skewness and kurtosis of IRAS galaxies in redshift
space. Our results agree well with previous measurements in both the
parent angular catalogue, and in the derived redshift surveys. This is
in contrast with smaller, optically selected surveys, were there is a
discrepancy between the redshift space and projected
measurements. Predictions from cold dark matter theory, obtained using
the recent semi-analytical model of galaxy formation of Benson {\it et
al}, provide an excellent description of our clustering data.

\end{abstract}


\section{Introduction}

The clustering pattern of galaxies observed today reflects an
interplay between two fundamental processes: the gravitational growth
of primordial density fluctuations and the physics of galaxy
formation. In general, we expect the process of galaxy formation to
segregate the galaxies from the underlying dark matter distribution
and even to give rise to dependencies of clustering on physical
properties, such as galaxy colour, luminosity, morphological type or
star formation rate (e.g., Kauffmann \etal 1999, Benson \etal
2000). Such segregation is usually referred to as ``biasing.''

Biasing was originally introduced as a useful device to match
theoretical predictions for the distribution of cold dark matter (CDM)
with observations of the spatial correlations of clusters and galaxies
(Kaiser 1984, Davis \etal 1985, Bardeen \etal 1986). Since then, the
statistical properties of the dark matter distribution in CDM models
has been extensively investigated by means of N-body simulations
(e.g. Jenkins \etal 1998, Gross \etal 1999). Observationally, however,
little is known about the large-scale distribution of dark matter,
although this situation may soon change as gravitational lensing
techniques become increasingly sensitive (Fischer \etal 2000). In the
meantime, partial information on biasing may be obtained by studying
the relative clustering of galaxies with different physical
characteristics. This, in turn, requires large surveys like the one we
analyze in this paper.

The simplest statistical tool to quantify clustering is the two-point
correlation function, defined as the excess probability above (or
below) random that an object be found at a certain separation from
another, randomly chosen object. Two-point statistics completely
describe a Gaussian point process. However, the galaxy distribution is
patently non-Gaussian, as evidenced, for example, by the presence of
rich clusters and walls. A natural generalization is the complete set
of $N$-point correlation functions which provide a full
characterization of a distribution.  Unfortunately, their measurement
and interpretation become exponentially difficult as the order, $N$,
increases. This is why counts-in-cells analysis, which extracts the
{\it average} of the $N$-point correlation functions over a cell, has
become the most practical and popular tool for describing higher order
clustering in the galaxy distribution (e.g., Peebles 1980, Efstathiou
\etal 1991, Bouchet \etal 1993, Szapudi, Meiksin, \& Nichol 1996,
Gazta\~naga 1994, Szapudi \& Szalay 1997, Szapudi, \& Gazta\~naga
1998).

This paper calculates two-point statistics and presents a
counts-in-cells analysis of the PSCz survey of IRAS galaxies (Saunders
\etal 2000). Our analysis is motivated, in part, by a desire to
characterize some of the possible manifestations of biasing, such as a
dependency of clustering on luminosity or a particular functional form
for high order statistics. To this aim, we construct volume-limited
catalogues from the PSCz survey which provide a controlled,
homogeneous framework for statistical analysis. The most efficient
tool to search for a luminosity dependence of clustering is the
luminosity mark correlation (Beisbart and Kerscher 2000) which we
apply to the PSCz.  We also compare our measurements with a model of
galaxy formation in the context of the CDM cosmology (Benson \etal
2000) which makes specific predictions for the statistics that we
investigate.  A previous search for luminosity effects in the
clustering of IRAS galaxies, based on the QDOT survey, failed to
reveal any signal (Moore \etal 1994). However, a preliminary analysis
of the PSCz survey (Maddox \etal 2000) suggests that some effect could
well be present.

The rest of this paper paper is organized as follows.  In \S2, we present
brief details of the dataset and describe our measurements of the two-point
correlation function, mark correlations, and the moments of counts-in-cells
up to fourth order. In \S3, we compare our estimates with measurements for
other surveys, both optical and infrared, and with the prediction of the
Benson \etal model.

\section{Dataset and Measurements}

In this section, we characterize the clustering of galaxies in the
PSCz redshift survey using three complementary descriptors: the
two-point correlation function (\S 2.2), the mark correlations (\S
2.3), and the higher order correlation amplitudes or cumulants, the
skewness and kurtosis (\S 2.4). All three statistics were calculated
for a series of volume-limited subsamples, each containing all PSCz
galaxies in spheres of radius $50, 75, 100, 125, 150, 175$, and
$200\hmpc$, namely $1160, 2159, 2189, 1985, 1671, 1456$, and $1259$
galaxies, respectively.  We chose to work with volume-limited
subsamples because these are homogeneous catalogues that are simple to
treat statistically and, as shown by Colombi, Szapudi and Szalay
(1998), the full series of catalogues yields results that are
equivalent to a measurement with optimal weights in the full
catalogue.

\subsection{The PSCz Catalogue}

PSCz is a redshift survey of all the IRAS galaxies in the Point Source
Catalog that have a flux at 60 $\mu m$ greater than 0.6 Jy.  With its
15,500 galaxies distributed over 84.1\% of the sky, it constitutes the
deepest and densest all-sky redshift survey to date. The median depth
is just 8100 km s$^{-1}$, although useful information is available out
to 30,000 km s$^{-1}$ at high galactic latitudes and to 15,000 km
s$^{-1}$ elsewhere.  Full details of the survey may be found in
Saunders \etal (2000). The analysis was restricted to the region of
sky covered by the survey, with no attempt to interpolate the galaxy
distribution within the masked areas (Branchini \etal 1999).
%full details of the masks applied are found in Branchini \etal (1999).   
Since the sample is relatively local, we can ignore evolutionary
effects on the clustering and biasing.
% IS (which may,
% however, be significant in the next generation redshift surveys, such as the
% 2DF and SDDS projects).

\subsection{Correlation Function}

We estimate the two-point correlation function of the PSCz survey by
means of the Landy and Szalay (1993; LS) estimator which has been
shown to be optimal (Szapudi \& Szalay 1998; Kerscher, Szapudi, \&
Szalay 1999). Denoting the number of data-data, data-random, and
random-random pairs in a bin at distance $r$ by $DD, DR$, and $RR$
respectively, the definition of the estimator is
\begin{equation}
  \hat\xi(r) = (DD -2 DR + RR)/RR.
\end{equation}
%This estimator can be shown to provide near optimal edge corrections.  
For volume-limited surveys, uniform weighting yields minimum variance.
% (see the references above for 
% further details.) 
%
%
\begin{figure}% [ptb]
\centerline{\hbox{\psfig{figure=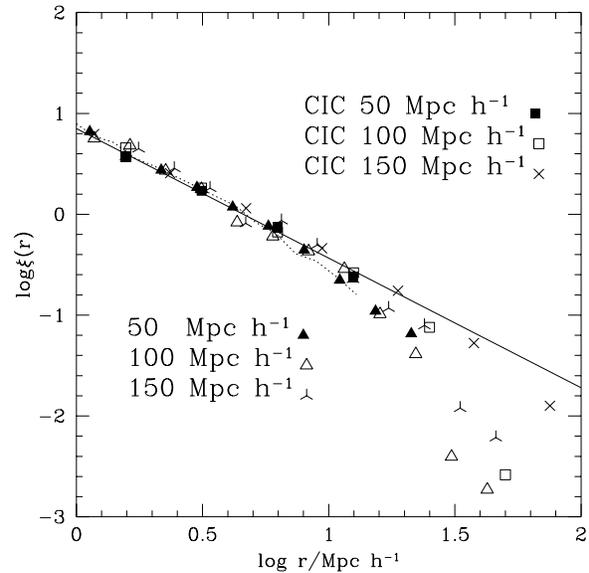,width=8cm}}}
\caption{The two-point correlation function of the PSCz survey
in redshift space. Three-sided symbols represent measurements using
the Landy-Szalay estimator, while four-sided symbols show estimates
derived from scaled counts-in-cells (c.f. \S2.4).  Results for three
representative volume-limited subsamples are displayed.  Statistical
errorbars are typically smaller than the symbols (except on the
largest scales), and are not plotted for clarity. The solid line is
the fit by Fisher \etal (1994, see also Seaborne \etal 1999). The
dotted line shows the predictions of the CDM model of Benson \etal
(2000) for spiral galaxies.}
\label{fig:figure1}
\end{figure}       

The three-sided symbols in Fig.~1 show our estimate for the PSCz
two-point correlation function using the Landy-Szalay estimator on
volume-limited samples of radii $50, 100$, and $150 \hmpc$. The square
symbols show a different estimate based on the counts-in-cells (CIC)
analysis described in \S2.4. While a rigorous error calculation for
the LS estimator is not available at present, errors for the CIC
estimator can be computed using the FORCE package (see \S2.4). The
1-$\sigma$ uncertainties are typically less than $5-7 \% $, but grow
rapidly at large separations, becoming $20-25\%$ at $r \sim 25 \hmpc$
and larger beyond this. The LS and CIC estimators are only expected to
agree in the regime where the correlation function is well
approximated by a power-law regime. In this regime the uncertainties
in both methods are most likely comparable. (The error bars have been
omitted from the figure for clarity.) It is important to recognize
that the data points in Fig.~1 (and in most other plots in this paper)
are not independent and so the estimated errors cannot be used for
maximum likelihood estimation without further analysis. This requires
computing the cross-correlation matrix of all measurements, a
complicated and ill-understood procedure which is beyond the scope of
this letter.

Results for the remaining volume-limited subsamples are consistent
with those displayed in the figure. For comparison, the solid line
shows the fit of Fisher \etal (1994) to the correlation function of
the IRAS $1.2$-Jy survey, $\xi(r) = (r/r_0)^\gamma$, with $r_0 = 4.43
\hmpc$, and $\gamma = 1.28$.  Our results are consistent with this
fit, perhaps with a cut-off on large scales, as well as with the fit
of Seaborne \etal (1999) to the full (flux-limited) PSCz survey.
Since the deeper samples contain predominantly brighter galaxies, the
similarity of the correlation functions in the subsamples suggests
that any luminosity dependence of clustering is small.  However, the
different samples have galaxies in common and so their correlation
functions are not independent. This weakens the sensitivity of this
test for a luminosity dependence and so we prefer, instead, to use the
mark correlations, as we now describe.

\subsection{Luminosity Mark Correlation Function}

The elegant tool of mark correlations, introduced into astrophysics by
Beisbart and Kerscher (2000), is ideal for quantifying the luminosity
dependence of clustering.
%The basic relevant definitions are summarized next.
%For more details, and a
%wider range of applications, we refer the reader to the KB. 
If $m_i$ denotes a mark, 
%which is identified with a luminosity cut in this
%paper, but which, in principle, could be any
or a parameter characterizing a class of objects, the joint
probability of finding a pair of galaxies at $(m_1,r_1)$ and
($m_2,r_2$), respectively is
\begin{equation}
  \Gamma(m_1,m_2,r_1, r_2)dm_1 dm_2 d^3r_1 d^3r_2 
\end{equation}
(see also Peebles 1980). If the mark is discrete, $dm_1$ is a
Stieltjes-Lebesgue measure; otherwise it is the usual Lebesgue
measure. The two-point correlation function is obtained by integration
over the marginal distribution,
\begin{equation}
  n^2\left(1+\xi(r)\right) = \int \Gamma dm_1 dm_2,
\end{equation}
where $r = |r_1-r_2|$.  The conditional probability of finding a pair
with marks $m_1$ and $m_2$ respectively is
\begin{equation}
  P(m_1,m_2 | r)  = \frac{\Gamma(m_1,m_2,r_1, r_2)}{ n^2(1+\xi(r))}.
\end{equation}
This quantity may be estimated from the catalogue as the ratio of all
the pairs with the prescribed marks divided by all the pairs, both
numbers taken at a given separation.
%IS According to KB, eqn~4 is an optimal
%estimator of the conditional probability. This is slightly counterintuitive,
%since edge corrections are the most significant issue when constructing
%estimators for spatial statistics. When calculating mark correlations,
%however, the corrections approximately cancel out because of the division
%by the full correlation function.  It is customary to take moments
%of the probability distribution, but for our purposes it will suffice to
%consider it in its pure form.

The luminosities were divided into luminosity quartiles, each of them
containing equal numbers of galaxies. Thus, marks $m = 0,1,2,3$
represent bins of increasing absolute luminosity.  The rest of this
subsection is exclusively concerned with the shallowest volume-limited
subsample, cut at a distance of $50\hmpc$, since this contains the
largest range of luminosities.  The same analysis was repeated for the
other volume-limited subsamples with similar results.

The probability, $P(m_1,m_2|r)$, is shown in Fig~2, as a function of $4
m_1+m_2, m_1 \ge m_2$ for a number of pair separations.
%, $ 2.35, 3.26, 4.51,
%6.25, 8.67, 12.01\hmpc$ (upper panel) and $16.64, 23.06, 31.94, 44.26,
%61.32, 84.95 \hmpc$ (lower panel).
%
%
\begin{figure}% [ptb]
\centerline{\hbox{\psfig{figure=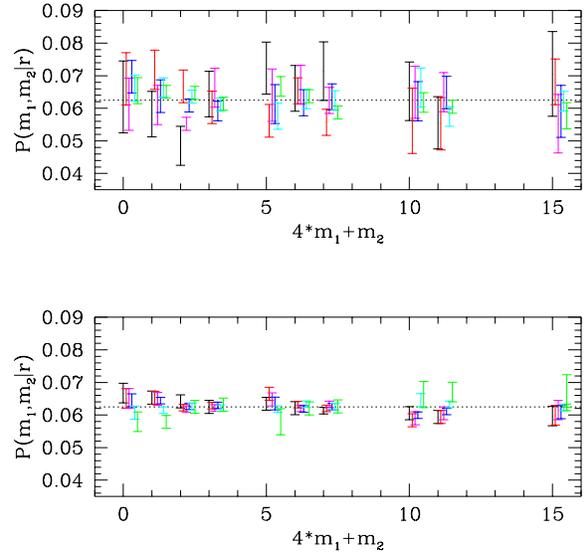,width=8cm}}}
\caption{The conditional probability $P(m_1,m_2|r)$. This is plotted 
as a function of $4m_1+m_2$ (with $m_1 \ge m_2$) in order to separate
the points.  Here, $m_i = 0,1,2,3$ represents the luminosity quartiles
of a volume-limited PSCz subsample with radius of $50\hmpc$, with the
corresponding luminosity ranges between 
%$(0.6, 0.783, 1.092, 1.85, 28.4)$ Jy.  
$(9.06\, 10^{42}, 1.18233\, 10^{43}, 1.64892\, 10^{43}, 2.7935\, 10^{43},
4.2884\, 10^{44})$ ergs/sec.
For clarity, only symmetrical errorbars, estimated from a series
of 10 realistic mock PSCz catalogues, are shown. Results for several
scales are plotted, with small horizontal shifts to the right (for
clarity), in increasing order: $2.35, 3.26, 4.51, 6.25, 8.67,
12.01\hmpc $ (upper panel) and $16.64, 23.06, 31.94, 44.26, 61.32,
84.95 \hmpc$ (lower panel).}
\label{fig:figure2}
\end{figure}       
For a particular pair ($m_1, m_2$), results corresponding to
increasing scales are slightly shifted towards the right for clarity
and only symmetric errorbars are plotted. These were calculated from
the dispersion of a set of 10 mock PSCz catalogues constructed from an
N-body simulation by Cole et al. (1998) of a flat $\Lambda$CDM model,
with cosmological constant, $\Lambda=0.7$, and without any luminosity
segregation.  Fluxes were assigned randomly to galaxies to match the
PSCz luminosity function and the various selection criteria discussed
by Branchini \etal (1999) were applied in order to obtain realistic
mock catalogues. For the smaller distances considered (upper panel),
the errors agree with a naive Poissonian estimate, while for the
larger scales (lower panel), the dispersion in the mock catalogues
significantly exceeds the Poissonian expectation.
%Although a rigorous calculation of the errors on
%$P(m_1,m_2|r)$ has not been attempted so far, it is clear that 
Therefore, on small scales the dominant source of error is discreteness or 
shot noise,  while on large scales finite volume effects are important. 
% IS Note that our analysis is
% not affected by edge effects since the estimator weights each pair
% uniformly.

The dotted line in Fig.~2 corresponds to the theoretical expectation
for the probability, $p = (1/4)^2= 0.0625$, for no correlation between
the bins. Points above or below this line indicate a positive or
negative correlation respectively.  The results in Fig.~2 are
consistent with no correlation on any scale, for all pairs of marks.
% about 10 points out of 50 is outside of the one sigma barrier.
%In the lower panel, there appears to be a certain degree of coherence
%between the points, hinting at a slight anticorrelation between bright
%bins and a positive correlation between faint bins which increases
%with scale, with no signal for intermediate bins. This effect could be
%attributed to a combination of a chance alignment with the
%normalisation condition for $P(m_1,m_2|r)$, whose sum over all bins
%must add up to unity. The same procedure returned a similar degree of
%coherence at least once in 10 mock surveys. 
To quantify this statement, we use the 10 mock catalogues just
mentioned.  The formal $\chi^2$ for all the points plotted, $\chi^2 =
0.381664$, would be unrealistically small for 120 independent
points. However, the points in the figure are not independent because
the normalization condition for $P(m_1,m_2|r)$ requires the sum over
all the bins to add up to unity.  For this reason, the formal $\chi^2$
has a non-standard distribution which can, in principle, be derived by
Monte-Carlo methods using our mock catalogues. Although 10 mock
catalogues are not sufficient to determine the distribution in detail,
they allow us to reject a significant signal in Fig.~2: the $\chi^2$
for the sample is close to the median for the simulations, $0.381914$,
whose highest and lowest 10 values are $0.204304$ and $0.512983$,
respectively.

Our results agree with those from a similar, but independent, analysis
of the PSCz survey by Kerscher and Beisbart (private communication).
However, they disagree with preliminary results by Maddox
\etal (in preparation) who, using different subsamples and statistical
techniques, find a small but significant dependence of clustering on
luminosity in the PSCz catalogue. Nevertheless, on the basis of
analysis performed here, we may assume that any luminosity effects are
sufficiently small that different volume-limited subsamples can be
combined together to obtain minimum variance estimates of the higher
order moments, as we do in the next section.

Before proceeding, we note that a comparative study of the IRAS 1.2 Jy
and PSCz surveys by Teodoro \etal (1999) shows that the two density
fields differ, within $\sim 80 \hmpc$, by a monopole term, a finding
which is consistent with the spherical harmonics analysis of the PSCz
survey by Tadros \etal (1999).  Both these studies show that this
apparent discrepancy disappears at a flux limit of 0.75 Jy. We have
therefore repeated our analysis using samples limited at 0.75 Jy. Our
results remain essentially unchanged for this sample, with a formal
$\chi^2=0.468862$ for the comparison in Fig.~2.

\subsection{Counts-in-Cells Analysis}

We carried out a counts-in-cells analysis following the prescription
developed by Szapudi \etal (1999). In a nutshell, we measured
counts-in-cells using the successive convolution algorithm described
in that paper, with a high rate of oversampling (up to $10^9$ cells)
on all scales for each volume-limited subsample.  In practice, only
about half of the cells intersected the geometry of the catalogue
embedded in a cubic grid. From the counts-in-cells, the skewness and
kurtosis were calculated using the method of Szapudi \& Szalay (1993).
This technique automatically corrects for shot noise by replacing
moments with factorial moments. Thus, continuous definitions of the
cumulants suffice:
%\begin{eqnarray}
\begin{equation}
%  \xib && = \avg{\delta^2} \nonumber \\ 
%  S_3 && = 3 Q_3 = \frac{\avg{\delta^3}}{\xib^2} \\
%  S_4 && = 16 Q_4 = \frac{\avg{\delta^4}}{\xib^3}, \nonumber
  S_N  \equiv N^{N-2} Q_N = \frac{\avg{\delta^N}}{\xib^{N-1}}, \\
%\end{eqnarray}
\end{equation}
where $\delta$ denotes the field of density fluctuations, and $\xib =
\avg{\delta^2}$ ($N = 3,4$ for this paper). In the power-law
approximation, the average correlation function is proportional to the
correlation function.  A Monte Carlo integration of the power-law fit
to $\xi(r)$ given in the previous subsection yields $\xib/2.4 =
\xi$. We plot $\xi(r)$ derived from the first of equations~(5) and
this relation in Figure~1 (filled squares). The result is consistent
with the direct estimate of $\xi(r)$ at pair separations at which the
power-law approximation is valid. For larger scales the above relation
is no longer valid.

We determined the $S_3$ and $S_4$ for all the volume-limited
subsamples and, for each cell-size, we cherry-picked the value with
the smallest error. The analysis of the preceding sections suggests
that any luminosity bias is small, so we can assume that the different
volume-limited subsamples represent the same statistical process.
\begin{figure}% [ptb]
\centerline{\hbox{\psfig{figure=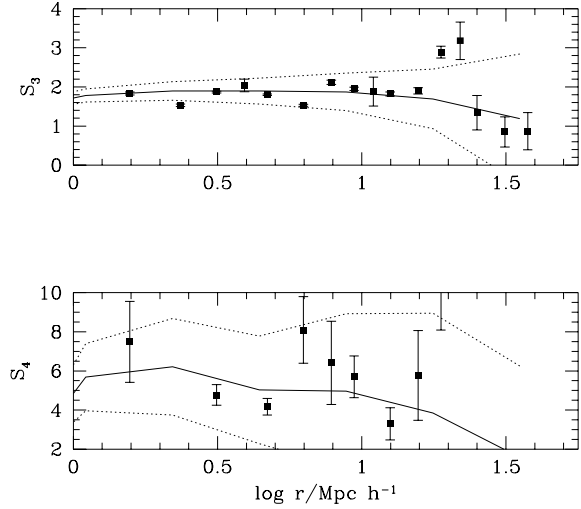,width=8cm}}}
\caption{Near optimal composite measurements of $S_3$ (upper panel) and $S_4$
(lower panel). Different points can come from different volume-limited
cuts, as detailed in Table~1. The errorbars were calculated using the
SS model.  The solid lines with the enveloping dotted lines
representing the variance are the predictions of the semi-analytic
model of galaxy formation of Benson \etal (1999) with the
corresponding variance.  }
\label{fig:figure3}
\end{figure}       
Colombi \etal (1998) have shown that in these circumstances,
selecting the highest signal-to-noise measurements from a series of
volume-limited surveys is equivalent to the optimal minimum variance
weighting of the flux-limited survey.

The errors were determined {\it ab initio} using the FORCE (FORtran
for Cosmic Errors) package based on the theory of Szapudi \& Colombi
(1996), and Szapudi, Colombi \& Bernardeau (1999).  For each
volume-limited subsample, the error was obtained for each order and
scale under a range of plausible assumptions for the parameters
involved. Whenever possible, these parameters were measured from the
survey itself, such as its volume, the average number-counts and the
variance in a cell.  The variance over the survey volume was derived
assuming the standard CDM power spectrum with $\Gamma = 0.5$ and
$\sigma_8 = 1$.  The higher order cumulants for $3 \le S_3
\le 8$ were calculated from perturbation theory with
the effective index of the power spectrum $n = -0.5$.  This yields
slightly higher $S_N$s than are typically measured in the PSCz survey,
thus providing a conservative overestimation of the (statistical)
errors.  Finally, the assumption of Szapudi \& Szalay (1993, SS) was
employed for the structure of cumulant correlators, the two point
analogs of the $Q_N$s, namely $Q_{NM} = Q_N Q_M$ (see SS for details).
%Alternatives, such
%as perturbation theory (PT, e.g. Juszkiewicz \etal
%1993, Bernardeau 1994) were examined as well, but the results
%appear to be robust against these variations.

Provided that the parameters are well tuned, the calculations
typically provide better than 50\% accuracy for the statistical
errors, normally a conservative overestimate as verified by Colombi
\etal (1999) and Hoyle \etal (1999) from N-body simulations.  It
should, however, be borne in mind, that if the plausible guesses made
for the parameters and models are grossly incorrect, or if systematic
errors dominate, the theoretical error estimates might be
inaccurate. We have varied some of the assumptions and parameter
values within reasonable bounds without finding significant change.
When the relative errors are approaching unity, the perturbative
approach breaks down, but still indicates that the measurement has low
significance.  For more details of the method and its applicability
see Szapudi \etal (1999).  
%Only points with relative errors less than about $60$\% were retained
%as an estimate of $S_3$ and $S_4$ from each volume-limited sample. For
%each scale, the highest signal-to-noise measurement was selected and
%displayed in Fig.~3 and Table~1.  
Note that these error estimates do not take into accoount
cross-correlations or possible systematic errors and could therefore
underestimate the true uncertainties.

\begin{table}
\caption{Estimates of $S_3$ in cubic cells of size $l$. The
errors, $\sigma_{S_N}$, were calculated using the FORCE package.  The
depth of the volume-limited subsample picked out by the near optimal
procedure (see text) is denoted by $R$ and given in $\hmpc$. The value
of $R$ corresponding to each estimate of $S_3$ is also applicable to
the corresponding estimate of $S_4$, except for the case indicated by
an $^*$, for which $R = 50\hmpc$ for $S_4$.}
\begin{tabular}{lccccc}
\hline
$l$ & $S_3$ & $\sigma_{S3}$ & $R$ & $S_4$ & $\sigma_{S_4}$ \\ \hline
0.78 & 2.77 & 0.35 & 50 & - & -     \\    
1.57 & 1.83 & 0.04 & 50 & 7.49 & 2.07 \\    
2.35 & 1.53 & 0.03 & 75 & - & -     \\    
3.14 & 1.87 & 0.02 & 50 & 4.77 & 0.53\\    
3.92 & 2.04 & 0.16 & 125  & - & - \\   
4.71 & 1.80 & 0.02 & 75  & 4.17 & 0.42 \\    
6.27 & 1.53 & 0.02 & 100 & 8.09$^*$ & 1.70  \\   
7.84 & 2.12 & 0.06 & 125 & 6.41 & 2.13  \\   
9.41 & 1.95 & 0.04 & 75  & 5.70 & 1.06  \\    
10.98 & 1.88 & 0.37 & 175  & - & - \\  
12.55 & 1.82 & 0.05 & 100 & 3.30 & 0.82   \\  
15.69 & 1.90 & 0.08 & 125 & 5.77 & 2.29   \\  
18.82 & 2.89 & 0.15 & 150 & 17.24 & 9.15  \\  
21.96 & 3.18 & 0.48 & 175  & - & - \\
25.10 & 1.34 & 0.44 & 100  & - & - \\  
31.37 & 0.85 & 0.38 & 125  & - & - \\  
37.65 & 0.87 & 0.48 & 150  & - & - \\  \hline
\end{tabular}
\end{table}
\section{Discussion and Summary}

We have characterized the clustering properties of galaxies in the
PSCz survey of IRAS galaxies by means of the two-point correlation
function, mark correlations, and the moments of counts-in-cells. Our
two estimates of the correlation function, one based on a direct
measurement and the other derived from counts-in-cells, are consistent
with each other and agree well with previous measurements for IRAS
galaxies (Moore \etal al 1994, Fisher \etal 1994, Seaborne \etal
1999). Neither the correlation function itself nor the mark
correlations show any significant evidence for a dependence of
clustering on luminosity, over the limited range of luminosities
probed by our volume-limited subsamples.  Finally, we have firmly
established that the skewness of the distribution of counts-in-cells
in redshift space for IRAS galaxies has a value $S_3 \simeq 2$.

Our estimated values of $S_N$, for $N = 3, 4$, are listed Table~1 and
displayed in Fig.~3. The variation of these quantities over the range
of scales probed by our analysis is small. Over the entire range, $S_3
= 1.89 \pm 0.62$ and $S_4 = 7.00 \pm 4.13$, while over the restricted
range $1-20\hmpc$, $S_3 = 1.93 \pm 0.35$. These results are in good
agreement with previous analyses of counts-in-cells of IRAS galaxies,
based on the QDOT survey (Saunders \etal 1993). For the parent
catalogue of the 1.2Jy survey, Meiksin \etal (1992) obtained $S_3 =
2.2 \pm 0.2$ and $S_4 = 10 \pm 3$, while for the 1.2 Jy redshift
survey itself Bouchet \etal (1993) found $S_3 = 1.5 \pm 0.5$ and $S_4
= 4.4 \pm 3.7$ and Fry \& Gazta\~naga (1994) derived $S_3 = 2.1
\pm 0.3$ and $S_4 = 7.5 \pm 2.1$.  The significance of our study 
lies in the more densely sampled data set, the use of state-of-the-art
measurement techniques, and a rigorous error calculation. All these
features have enabled us to extend the dynamic range of previous
studies and achieve unprecedented accuracy.

In contrast to the situation at optical wavelengths (see Hoyle
\etal 1999 for an up-to-date discussion), the skewness of
IRAS galaxies in redshift space agrees well with that in the parent,
projected catalogue, as measured by Meiksin \etal (1992). Presently,
these are the only datasets large enough to allow a measurement of
$S_3$ with a precision better than 10\% over a large dynamic range.
The 2dF and SDSS surveys will enable estimates of $S_3$ with an error
of only a few percent and of $S_4$ with an error of 10-25\% over a
large dynamic range (Szapudi \etal 1999).  These surveys will clarify
the role of redshift space and projection effects in the apparent
disagreement between angular and 3D cumulants for optically selected
galaxies, and enable a more direct comparison between the statistics
of IRAS and optical galaxies.

In general, the flatness of the $S_3$ curve in redshift space is well
understood from theoretical arguments, and supported by $N$-body
simulations. Although the skewness rises to a non-linear plateau in real
space, the random velocities associated with the 'finger of god' distortion
in redshift space act as an effective smoothing.  As a result, the $S_3$
curve remains flat in redshift space.  Naturally, biasing complicates this
picture inferred from simple theory and dark matter simulations. It is
therefore perhaps surprising that $S_3$ should be so similar for IRAS and
optically-selected samples since these have somewhat different spatial
distributions. We speculate that this agreement is the result of a
cancellation effect whereby the differences in spatial distribution are
compensated for by differences in the strength of the redshift space
distortions acting on each galaxy type.

To address the effects of biasing on statistical measurements of the
galaxy distributions requires a detailed theory of galaxy formation.
The solid line in Fig.~3. shows the predictions of the semi-analytic
model of galaxy formation proposed by Cole \etal (2000) and Benson
\etal (1999). According to the model, the higher order moments in {\it
redshift space} for spiral galaxies are very similar to those for the
galaxy population as a whole, although they differ in real space
(Baugh, Szapudi, \& Benson 2000).  The predictions in the figure refer
to spiral galaxies which, as a class, are a reasonable representation
of IRAS galaxies.  The dotted lines give the variance computed by
varying the input parameters in the error estimation procedure.  The
agreement with the PSCz results is remarkably good. The shape of the
$S_3$ and $S_4$ curves in the model reflects the kind of cancellation
effects mentioned above (see Baugh \etal 2000 and Hoyle \etal 1999 for
further discussion). As shown in Fig.~1, the model predictions also
agree well with the two-point correlation function of the PSCz
survey. This model of galaxy formation assumes a flat,
$\Lambda$-dominated CDM cosmology in which galaxies form by
hierarchical clustering. The agreement with our results is the more
remarkable since there are no adjustable parameters in the comparison,
so that the theoretical lines in Figs.~1 and ~3 are to be regarded as
genuine predictions of the model.

%\begin{figure}
%\centering
%\centerline{\epsfysize=9.truecm
%\epsfbox{s3.ps}}
%\caption[junk]{Same as in Figure \ref{sigma} for the 
%the hierarchical skewness $s_3=w_3/w_2^2$.
%The misalignment of the open and solid  squares on scales $gt 0.5$ degrees
%is the  result of edge effects, as both correspond to smaller surveys.}
%\label{s3}
%\end{figure}

\bigskip

{\bf Acknowledgments}

We thank the rest of the PSCz team for use of the data before
publication, S. Cole for providing mock PSCz catalogues and A. Benson
and C. Baugh for providing theoretical predictions from the Durham
semi-analytic model of galaxy formation and for useful comments.  IS
was supported by the PPARC rolling grant for Extragalactic Astronomy
and Cosmology at Durham. CSF acknowledges a Leverhulme Research
Fellowship. The FORCE (FORtran for Cosmic Errors) package is available
from its authors, S. Colombi and IS
(http://www.cita.utoronto.ca/$^{\sim}$szapudi/istvan.html).

\def\aj { AJ}
\def\apj { ApJ}
\def\aap {A \& A}
\def\ajs{ ApJS}
\def\apjs{ ApJS}
\def\mnras { MNRAS}
\def\apjl { Ap. J. Let.}


\begin{thebibliography}{}

\bibitem[Benson \etal 1999a]{bbcfl99} Benson, A., Baugh, C., Cole, S.,
Frenk, C.S. \& Lacey, C. 1999, \mnras, submitted (astro-ph/9910488)
\bibitem[Baugh, Szapudi \& Benson 2000]{bsb00} Baugh, C.M.,
Szapudi, I., Benson, A. 2000, in preparation
\bibitem[Beisbart & Kerscher 2000]{bk00}  Beisbart, C, Kerscher, M
2000, \apj, submitted (astro-ph/0001036)
\bibitem[Bernardeau 1994]{bern94}  Bernardeau, F. 1994, \apj, 433, 1
\bibitem[Branchini \etal 1999]{pscz2}Branchini E., \etal
%L. Teodoro, C.S. Frenk, I. Schmoldt, G. Efstathiou, S.D.M. White, W. Saunders,
%W. Sutherland, M. Rowan-Robinson,   O. Keeble, H. Tadros, S. Maddox, 
%S. Oliver, 
1999, \mnras, 308, 1
\bibitem[Bouchet \etal 1993]{bouchet93}  Bouchet, F.R., \etal
%         Strauss, M.A., Davis, M., 
%         Fisher, K.B., Yahil, A., \& Huchra, J.P. 
1993, ApJ, 417, 36
\bibitem[Cole, \etal 1998]{chwf} Cole, S., Hatton, S., Weinberg, D. \&
Frenk, C.S. 1998, \mnras, 300, 945
\bibitem[Cole, \etal 1999]{chwf} Cole, S., Lacey, C., Baugh, C. \& 
Frenk, C.S. 1999, \mnras, submitted. 
\bibitem[Colombi, Szapudi, \& Szalay 1998]{css98}  Colombi, S., Szapudi, I., 
Szalay, A.S., 1998, \mnras, accepted (astro-ph/9711087) 
\bibitem[Efstathiou \etal 1990]{e90}Efstathiou, G., \etal
% Kaiser, N.,
% Saunders, W., Lawrence, A., Rowan-Robinson, M. Ellis, R.S., Frenk, C.S.,
1990, \mnras, 247, 10 
\bibitem[Fischer \etal (1999)]{f99} Fischer \etal 1999
        \apj, submitted(astro-ph/9912119)
\bibitem[Fisher \etal (1994)]{f94} Fisher, K.B., Davis, M., Strauss, M., 
Yahil, A., Huchra, J., 1994, \mnras, 266, 50
\bibitem[Fry \& Gazta\~naga (1994)]{fg94} Fry, J.~N. \& Gazta\~naga, E. 1994,
        \apj, 425, 1        
\bibitem[Hoyle, Szapudi, \& Baugh 1999]{hsb99} Hoyle, F.,
        Szapudi, I., \& Baugh, C.M. 1999, submitted (astro-ph/9911351)
\bibitem[Juszkiewicz, Bouchet, \& Colombi 1993]{jbc93} Juszkiewicz, R.,
        Bouchet, F.~R., \& Colombi, S. 1993, \apj, 412, L9
\bibitem[Kerscher, Szapudi \& Szalay 1999]{kss} Kerscher, M., Szapudi, I. \& 
Szalay, A. 1999, submitted (astro-ph/9912088)
\bibitem[LS]{ls93}  Landy, S.D.,\& Szalay, A.
         1993, \apj, 412, 64 
\bibitem[Meiksin, Szapudi, \& Szalay 1992]{mss92}  
         Meiksin, A., Szapudi, I., \& Szalay, A., 
         1992, \apj, 394, 87 
\bibitem[Moore \etal 1994]{moore94}Moore, B., Frenk, C.S., Efstathiou, G. 
\& Saunders, W., 1994, \mnras, 269, 742
\bibitem[Saunders \etal 2000]{pscz1} Saunders \etal 2000, \mnras,
(submitted astro-ph/0001117)
%\bibitem[Schmoldt \etal 1999]{pscz3} Schmoldt \etal 1999, \aj, 118, 1146
\bibitem[Seaborne \etal 1999]{pscz4} Seaborne \etal 1999, \mnras,
(accepted astro-ph/9905182)
\bibitem[SC96]{sc96} Szapudi, I., \& Colombi, S. 1996,
        \apj, 470, 131 
\bibitem[Szapudi, Colombi, \& Bernardeau 1999]{cbs} Szapudi, I., Colombi, S.,
Bernardeau, F., 1999, \mnras, accepted 
\bibitem[Szapudi \& Szalay 1993a]{ss93} Szapudi, I. \& 
Szalay, A. 1993, \apj, 408, 43 (SS)
\bibitem[Szapudi \& Szalay 1998]{ss98} Szapudi, I. \& 
Szalay, A. 1998, \apj, 494, 41L
\bibitem[Tadros \etal 1999]{t99} Tadros, H. \etal 1999
\mnras, 305, 527
\bibitem[Teodoro \etal 1999]{t00} Teodoro, L. \etal 1999 
\mnras, submitted, (astro-ph/9908358).

\end{thebibliography}
\end{document}